\documentclass[pre, reprint, superscriptaddress]{revtex4-2}

\usepackage{amssymb}
\usepackage{amsmath}
\usepackage{amsfonts}
\usepackage[colorlinks,linkcolor=red,urlcolor=blue,citecolor=blue]{hyperref}
\usepackage{graphics}
\usepackage{txfonts}
\usepackage{tikz}  
\usetikzlibrary{arrows,shapes,positioning,shadows,backgrounds,fit}
\usepackage{braket}

\graphicspath{{figs/}}
\begin{document}

\title{Thermodynamics of light emission}
\author{Antoine Rignon-Bret}
\email{arignonbret@gmail.com}
\affiliation{\'{E}cole Normale Sup\'{e}rieure, 45 rue d'Ulm, F-75230 Paris, France}
\begin{abstract}
Some interactions between classical or quantum fields and matter are known to be irreversible processes. Here we associate an entropy to the electromagnetic field from well-known notions of statistical quantum mechanics, in particular the notion of diagonal entropy. We base our work on the study of spontaneous emission and light diffusion. We obtain a quantity which allows to quantify irreversibility for a quantum and classical description of the electromagnetic field, that we can study and interpret from a thermodynamical point of view. 
\end{abstract}

\maketitle

\section{Introduction}

 Quantification of irreversibility has been a main issue for a long time in physics, and becomes even more important nowadays with the development of new branches as quantum information or quantum thermodynamics. Among all the physical phenomena involving irreversible processes, we will focus here on some well known process appearing in field theory. The equations governing classical or quantum field theories are known to be reversible, however some processes described by these theories exhibit an irreversible behavior. For instance, we can think about radiation damping in classical electrodynamics \cite{feynman1965feynman}, emission of gravitational waves in relativistic gravitation \cite{einstein1937gravitational} or spontaneous emission in quantum electrodynamics \cite{scully1999quantum}. The way of this irreversible behavior arises from reversible field equations is very well understood and belongs to the established knowledge of physics. However, in this paper, we will try to quantify explicitly the degree of irreversibility of such matter-radiation interactions, using the framework of thermodynamics and statistical mechanics. We will focus here on the electromagnetic field and we will show that we can associate an entropy variation to such processes, which vanishes when there is no asymmetry between emission of light and absorption, but which is strictly positive in the opposite case.
 
The situation is analog to the fall of a ball in a gravity field. The ball falls because of gravity, but if there were no dissipation, it could bounce back and reach the initial height. However, it does not, because by hitting the ground, some part of the mechanical energy of the ball go to the atoms in the ground. They get excited and so the final state of the ball is the state of maximal entropy or minimal energy. If we had a precise enough thermometer, we could observe a temperature variation because of this excitation, and this is the signature of an increasing entropy variation. We aim to think some matter-radiations interactions, as the spontaneous emission process, in the same way, which means that we have to find the corresponding vibrating degrees of freedom which makes the ground state stable unlike the excited states, and get an entropy from them. By doing this, we will associate some entropy to the electromagnetic field. 
 
 The idea of associating an entropy to the electromagnetic field is old, as old as thermodynamics and electromagnetism are. Actually it is by confronting these two theories together that Einstein understood that radiation was made of indivisible quanta of energy \cite{einstein1905heuristic}, which lead to quantum mechanics, with the success we know. However, if Einstein worked with a thermostatted box where radiation was trapped and reached equilibrium with matter, the aim in the following work is different. We focus on quantum or classical electromagnetic processes, possibly involving a single atom and a single photon, so there is no associated thermal equilibrium and we are not in the thermodynamic limit. However, as entropy is an extensive quantity, unlike temperature, the notion still makes sense as long as we can count states. 
 
Finding the "good" microscopic definition of entropy is still a challenge. Aside from the well known Von Neumann entropy $S = - k_B Tr\big( \hat{\rho} ln \hat{\rho} \big)$, more general frameworks have been developed \cite{vsafranek2020brief}. For this work, we will be particularly interested in the diagonal entropy~\cite{polkovnikov2011microscopic}, which has been introduced to quantify the degree of irreversibility of some transformations in an isolated system. Indeed, the classical Von Neumann entropy associated to quantum systems always vanishes for unitary evolutions. In the following, after reminding the main features of spontaneous emission in section \ref{sec:spontaneousemission}, we will give in section \ref{electromagnetic field entropy} an appropriate definition of the entropy of the electromagnetic field, that we will relate to diagonal entropy. In particular, we will give a thermodynamical interpretation to the formula and we will check that it verifies the basic requirements for an entropy. In section \ref{classical electrodynamics}, we will talk a bit about classical electrodynamics and we will show that the electromagnetic field entropy formula obtained in the previous sections is still valid and makes sense. 

\section{The framework of spontaneous emission}
\label{sec:spontaneousemission}

To fully understand our purpose, it may be useful to return to the general framework of quantum electrodynamics and spontaneous emission. We will only remind some basic results and make some comments,  further details or proofs are given in Scully's book \cite{scully1999quantum}. The general hamiltonian is given by :

\begin{equation}
    \hat{H} = \hat{H}_{atom} +\hat{H}_{em} + \hat{H}_{int}
    \label{hamiltonientotal}
\end{equation}

\begin{equation}
    \hat{H} = \frac{\hbar \omega_0}{2}(Id + \hat{\sigma}_z) + \sum_{(\textbf{k}, s)} \hbar \omega_{\textbf{k}} \hat{a}_{\textbf{k}, s}^{\dagger}\hat{a}_{\textbf{k}, s} + 
    \sum_{(\textbf{k}, s)} \hbar \lambda_{\textbf{k}, s}(\hat{\sigma}_+\hat{a}_{\textbf{k}, s}+\hat{\sigma}_-\hat{a}_{\textbf{k}, s}^{\dagger})
    \label{hamiltonien}
\end{equation}

where :

\begin{equation}
    \lambda_{\textbf{k}, s} = \textbf{d} \cdot \textbf{u}_{\textbf{k},s}
    \sqrt{\frac{\hbar\omega_{\textbf{k}}}{2\epsilon_0V}}
    \label{lambdaks}
\end{equation}

is a result obtained from quantum field theory. Here, $\textbf{d}$ is the dipole moment operator matrix element between the atom ground state and excited state, $\textbf{u}_{\textbf{k},s}$ is a unit vector in Fourier space in the direction $\textbf{k}$ and polarization $s$, $\hat{a}_{\textbf{k}, s}$($\hat{a}^{\dagger}_{\textbf{k}, s}$) being the electromagnetic field state $(\textbf{k}, s)$ annihilation (creation) operator, $\hat{\sigma}_{+}$($\hat{\sigma}_{-}$) the atom quantum energy creation (annihilation) operator, and finally $V = L^3$ the volume of the cavity which encloses the atom. As we study the desexcitation of an atom, there is at most one photon in the cavity, and the state of the system atom + field is the following :

\begin{equation}
    \ket{\Psi(t)} = c_0(t) \ket{e}\ket{0} + \sum_{(\textbf{k}, s)} c_{(\textbf{k}, s)}(t) \ket{g}\ket{\textbf{k}, s}
    \label{etat du system}
\end{equation}

where $\ket{e}$($\ket{g}$) in the atom excited (ground) state. By using Schrodinger equation with hamiltonian \eqref{hamiltonien} and projecting it onto proper states we get the following equation system :

\begin{align}
    \dot{c}_0(t) = -i\omega_0c_0(t) - i\sum_{(\textbf{k}, s)} \lambda_{\textbf{k}, s}c_{\textbf{k}, s}(t)\\
    \dot{c}_{\textbf{k}, s}(t) = -i\omega_{\textbf{k}}c_{\textbf{k}, s}(t) - i \lambda_{\textbf{k}, s}c_0(t)
    \label{equationscouplées}
\end{align}

Here is the point. Until now, we considered that the atom was enclosed in a box of finite volume $V = L^3$. The size of the box is relevant because it constrains the electromagnetic modes that can propagate in the box by the quantification of the wave vector $\textbf{k}$. To lead the calculation further, we will assume that $L$ is very big such that the sums on $\textbf{k}$ can be replaced by integrals. Then, after using the Weiskoppf-Wigner approximation, we can show that :

\begin{equation}
    {c}_0(t) =  e^{-\frac{\Gamma}{2}t - i\omega_0t}
    \label{c0}
\end{equation}

With :

\begin{equation}
    \Gamma = \frac{d^2 \omega_0^3}{3\pi\hbar \epsilon_0 c^3}
    \label{gamma}
\end{equation}

From \eqref{c0}, the probability to find the atom in its excited state after time $t$ is simply :

\begin{equation}
    P_e(t) = e^{-\Gamma t}
    \label{probaexcité}
\end{equation}

The process is here clearly irreversible, the atom will stay forever in its ground state after de-excitation. The reason why we get an irreversible behavior starting from the Schrodinger equation which is clearly reversible is that we considered that the box was very big. If the box is relatively small, after emission, the photon can reflect on the cavity mirrors and come back to the atom in order to be reabsorbed, so with finite $L$, we do not get a decaying exponential but a periodic function, the photon being emitted, reflected, reabsorbed, then re-emitted, etc. If $L$ goes to infinity, the photon is never reflected by the walls of the cavity and the process is irreversible. Mathematically, a discrete sum of periodic function is still a periodic function, while a continuous sum (integral) of periodic function is not a periodic function anymore, so by doing the approximation of replacing sums by integrals we exhibited the irreversibility in the calculations.

Now, we are interested in the electromagnetic wave packet components, so in the values of the $c_{k,s}$. We can inject \eqref{c0} into \eqref{equationscouplées} to find (when $t>>\Gamma$) :

\begin{equation}
    c_{\textbf{k},s} = \frac{\lambda_{\textbf{k},s}}{(\omega_k-\omega_0)+i\frac{\Gamma}{2}}e^{-i\omega_{\textbf{k}}t}
    \label{cks}
\end{equation}

By taking the norm squared and by summing on the $\textbf{k}$ of constant modulus, we get the $\omega$ probability distribution :

\begin{equation}
    P(\omega) = \sum_{(\textbf{k},s)} \lvert c_{\textbf{k},s} \rvert^2 \delta(\omega_{k}-\omega) =  \frac{1}{\pi} \frac{\frac{\Gamma}{2}}{(\omega-\omega_0)^2+\frac{\Gamma^2}{4}}
    \label{probafreq}
\end{equation}

It is a lorentzian distribution, with standard deviation $\Delta \omega = \frac{\Gamma}{2}$ which is also its frequency width. Its time length is $\frac{1}{\Gamma}$, exactly as is the atom excited state lifetime. 

\section{Electromagnetic field entropy}

\label{electromagnetic field entropy}

\subsection{Statistical point of view}

As Eq.\eqref{etat du system} shows, the system atom + photon is always in a pure state. The matrix density of the system reads :

\begin{equation}
    \hat{\rho} = \ket{\Psi(t)} \bra{\Psi(t)} = \hat{U}_{\hat{H}}(t,0)\ket{\Psi(0)} \bra{\Psi(0)} \hat{U}^{\dagger}_{\hat{H}}(t,0)
\end{equation}

where $\hat{U}_{\hat{H}}(t,0)$ is the unitary evolution operator between times $0$ and $t$. The Von Neumann entropy $S = - k_B Tr\big( \hat{\rho} ln \hat{\rho} \big)$ always vanishes at all time because $\hat{\rho}$ is always in a pure state and it seems that we can't describe the irreversibility of the spontaneous emission process by a corresponding amount of entropy. It is true as long as we stay with the pure Von Neumann entropy. However, we can introduce a little trick. The system atom + electromagnetic field is not completely isolated, because there is a reflecting cavity which surrounds it. This cavity itself is placed in a dynamic universe and interact with it. Indeed, because of long-range interaction as gravity that we can never really remove, such tiny effects will play a role if we wait a long time enough. For instance, as the photon wave vector $k$ is related to its impulsion, and when the wave packet reflects on the cavity walls, it transfers some momentum to the box, that we can schematically write as :

\begin{align}
    \ket{\Psi(t)} = \bigg( c_0(t) \ket{e}\ket{0} + \sum_{(\textbf{k}, s)} c_{(\textbf{k}, s)}(t) \ket{g}\ket{\textbf{k}, s} \bigg) \otimes \ket{box} \longrightarrow  \\
    c_0(t) \ket{e}\ket{0}\otimes \ket{box} + \sum_{(\textbf{k}, s)} c_{(\textbf{k}, s)}(t) \ket{g}\ket{\textbf{-k}, s}\otimes\ket{2\textbf{k}}_{box}
\end{align}

By momentum conservation. Afterwards, the box can entangle with, for instance, dust present in the environment which starts a Von Neumann infinite regress \cite{laloe2001we}. If we are interested only in the system atom + electromagnetic field, we trace out the environment comprised of the box, dust, and any exterior system in the universe which interacts with our box. We get from \eqref{etat du system}:

\begin{align}
    \hat{\rho}_{sys} = Tr_{env}\hat{\rho}_{sys + env} \\
    = \lvert c_0(t) \rvert^2 \ket{e}\ket{0} \bra{0}\bra{e} + \sum_{(\textbf{k},s)} \lvert c_{\textbf{k},s} \rvert^2 \ket{g}\ket{\textbf{k}, s} \bra{\textbf{k}, s}\bra{g} 
    \label{mixed state}
\end{align}

Which is a diagonal density matrix representing a mixed state, and in consequence has non-vanishing Von Neumann entropy. We used here the framework of quantum decoherence \cite{laloe2001we,zurek2003decoherence}, which ensures that after a long enough time, all the off diagonal entries of the density matrix vanish, because of entanglement with an always existing environment. Likewise, diagonal entropy has been introduced by A.Polkovnikov \cite{polkovnikov2011microscopic} in order to study formally the irreversibility processes in quantum systems which are meant to be closed. The definition of this diagonal entropy is :

\begin{equation}
    S_d = - \sum_{n} \rho_{nn} ln\rho_{nn}
\end{equation}

where $\rho_{nn}$ is the $nn$-entry of the density matrix. Polkovnikov showed that it had all the good properties that we can expect from a suitable definition of entropy. This diagonal entropy has already been investigated by many authors \cite{santos2011entropy, giraud2016average, piroli2017correlations, wang2020diagonal, sun2020characterizing}. We can directly show that the diagonal entropy of the pure state \eqref{etat du system} is the von Neumann entropy of the density matrix \eqref{mixed state} obtained from a decoherent process. It reads :

\begin{equation}
    S = - \lvert c_{0} \rvert^2 ln\lvert c_{0} \rvert^2 -  \sum_{(\textbf{k},s)} \lvert c_{\textbf{k},s} \rvert^2 ln \lvert c_{\textbf{k},s} \rvert^2
    \label{S diagonal}
\end{equation}

As from \eqref{c0} $\lvert c_0 (t) \rvert$ goes from $1$ to $0$ when $t$ goes from $0$ to infinity, we can define the spontaneous emission entropy variation as $t \longrightarrow \infty$ as:

\begin{equation}
    \Delta S = - \sum_{(\textbf{k},s)} \lvert c_{\textbf{k},s} \rvert^2 ln \lvert c_{\textbf{k},s} \rvert^2
    \label{deltaS diagonal}
\end{equation}

which is a strictly positive quantity. 

\subsection{Explicit calculation of entropy}

We can calculate \eqref{S diagonal} and \eqref{deltaS diagonal} directly from the values of \eqref{cks}. However, we will make an approximation here to get a simpler result. Indeed, first, from \eqref{lambdaks} we know that $\lambda_{k,s}$ is proportional to $sin \theta$, where $\theta$ is the angle between the dipole moment vector $\textbf{d}$ and the wave vector $\textbf{k}$. This gives us that, typically, $\textbf{k}$ has non-zero probability distribution in the following solid angle (for a given norm of the wave vector $k_0$) :

\begin{equation}
    \Omega_{k_0} = \int_{0}^{\pi}\int_{0}^{2\pi} k_0^2 sin^3\theta \mathrm{d}\theta \mathrm{d}\phi = \frac{8 \pi \omega_0^2}{3c^2}
    \label{Volumefourier}
\end{equation}

The first approximation that we will make is setting that the distribution of the wave vector $\textbf{k}$ is uniform in this angular distribution $\Omega_{k_0}$ in Fourier space, and vanishes elsewhere Secondly, from \eqref{probafreq}, the frequency distribution is lorentzian. As we saw, the characteristic size of the frequency width is $\frac{\Gamma}{2} << \omega_0$ which is the lorentzian mean for standard values of the two quantities. So we will replace the lorentzian distribution by a uniform distribution centered on $\omega_0$ and of width $\delta \omega = \frac{\Gamma}{2}$. Thus, we can replace $ \sum_{(\textbf{k},s)} \longrightarrow \int \frac{\mathrm{d^3}k}{(\frac{2\pi}{L})^3}$  in \eqref{deltaS diagonal} and get from it :

\begin{equation}
    \Delta S \simeq \int_{- \infty}^{+ \infty} \frac{1}{\pi} \frac{\frac{\Gamma}{2}}{(\omega-\omega_0)^2+\frac{\Gamma^2}{4}} ln\bigg( \frac{8 \pi \omega_0^2}{3c^3} \bigg( \frac{L}{2\pi} \bigg)^3 \pi \frac{(\omega-\omega_0)^2+\frac{\Gamma^2}{4}}{\frac{\Gamma}{2}} \bigg) {\mathrm{d}\omega}
\end{equation}

\begin{equation}
    \Delta S \simeq ln \frac{V\omega_0^2\delta \omega}{3\pi c^3}
    \label{entropie3DFourier}
\end{equation}

The volume $K_0 = \frac{\omega_0^2\delta \omega}{3\pi c^3}$ is the typical volume in Fourier space of the available Fourier modes of the photon emitted. The more they are, the more "irreversible" is the desexcitation. We see that it corresponds to a volume $V_0$ in the real space such as :

\begin{equation}
    V_0 = \frac{3\pi c^3}{\omega_0^2\delta \omega}
    \label{V0}
\end{equation}

And from now we will write :

\begin{equation}
    \Delta S = ln\frac{V}{V_0}
    \label{entropie thermodynamique 1}
\end{equation}

The interaction between the electromagnetic field and the atom broadens the frequency range in Fourier space available for the photon. Irreversibility comes from the fact that there is not only one mode $\omega_0$ of the electromagnetic field that can be excited but many of them. However, the formula \eqref{entropie thermodynamique 1} gives us another interpretation. To make it clearer, let consider first the 1D case. In the same way as we derived \eqref{entropie3DFourier}, we can show that for a one-dimensional cavity we can associate to the electromagnetic field the entropy :

\begin{equation}
    \Delta S^{(1D)} = ln\frac{L\delta \omega}{2 \pi c}
    \label{entropie1Dfourier}
\end{equation}

Where $\delta \omega$ is the standard deviation of the one dimensional lorentzian frequency distribution. The wave packet typical length $\delta x$ can be obtained from the Heisenberg relation $\delta x \delta k \simeq \frac{1}{2}$, so we can write \eqref{entropie1Dfourier} as :

\begin{equation}
    \Delta S^{(1D)} \simeq ln\frac{L}{\delta x}
    \label{entropie1Dreal}
\end{equation}

Up to a irrelevant $4 \pi$ factor. Actually, it is true as long as we can set $c_0 = 0$ (remember that it is rigorously true only for $L \longrightarrow \infty$). In order to understand why, let consider the the typical time of de-excitation $\tau_{em}$, which is the inverse of the spontaneous emission rate. The typical time taken by the photon to explore the whole box is $\frac{L}{c}$. Furthermore, if we define $\tau_{ps}$ as the time taken by the energy quantum to explore the \textit{whole} phase space, we should consider the time $\frac{L}{c}$ when the energy quantum is propagating freely in the box and the time $\tau_{em}$, when the energy quantum is inside the atom (when the atom is in the excited state). So :

\begin{equation}
    \tau_{ps} = \frac{L}{c} + \tau_{em} 
    \label{temps espace des phases}
\end{equation}

as long as $L >> \delta x$, we can neglect $\tau_{em}$ before $\frac{L}{c}$ and the entropy is indeed given by \eqref{entropie1Dreal}. But in general we expect that the entropy is given by :

\begin{equation}
    \Delta S^{(1D)} = ln\frac{\tau_{ps}}{\tau_{em}}
    \label{entropie1time}
\end{equation}

because the entropy counts the number of states accessible to the energy quantum, and the time that the atom spends in the excited state $\tau_{em}$ is the same time as the wave packet spends in the mesh of size $\delta x \simeq c \tau_{em}$. The formula \eqref{entropie1time} enhances the fact that irreversibility is just a matter of time scale. 

Now, if we decide to decrease the size of the cavity up to it becomes on the same order as the typical length of the wave packet, the coefficient $c_0$ cannot be neglected anymore. In this case we get from \eqref{temps espace des phases} :

\begin{equation}
    \tau_{ps} \simeq \frac{L}{c} + \tau_{em} \simeq 2 \tau_{em}
    \label{temps espace des phases 2}
\end{equation}

Therefore, when $L \simeq \delta x$ :

\begin{equation}
    \Delta S^{(1D)} = ln\frac{2 \tau_{em}}{\tau_{em}} = ln2
\end{equation}

In that case, the atom will be half of the time in the excited state and half of the time in the ground state, and the process will almost seem to be reversible. We can make an analogy with a one particule Joule Gay-Lussac expansion, initially contained in a box of length $\frac{L}{2}$ and which can propagate in the whole box of size $L$ when the constraint is released. 

\vspace{0.3 cm}

In the three-dimensional case, the volume $V_0$ given by \eqref{V0} is not the typical volume of the wave packet, which is approximatly, for  $r = ct >> \frac{c}{\delta \omega} >> \frac{c}{\omega_0} $ :

\begin{equation}
    V_{\textbf{r}} \simeq \frac{8\pi}{3} r^2 \frac{1}{2\delta k}
    \label{Volumefonctiondonde}
\end{equation}

However, our theory tells us that if we wait a long time compared to the decoherence time, everything happens as if the wavepacket occupies a volume $V_0$ during a time $\tau_{em} = \frac{1}{\Gamma} = \frac{1}{2 \delta \omega}$. In other words, our phase space, which is the total volume $V$ is discretized in a mesh of volumes $V_0$, and the wave packet explores it. Actually, in this framework, the wave packet looks more like a particule than a wave packet, and the formula \eqref{entropie thermodynamique 1} is exactly the ideal gas entropy for one particule. If we add more excited atoms in the box (with the same energy gap), we should get :

\begin{equation}
   S = N ln\frac{V}{V_0}
    \label{entropie thermodynamique}
\end{equation}

as the photons do not interact. The philosophy here is we can see the spontaneous emission here as a purely thermodynamic process. Let suppose that at the beginning we "turned-off" the interaction between the atom and the electromagnetic field. The atom is forced to stay in the excited state and there is no entropy. At $t=0$, we turn on the interaction and the energy quantum can now move freely in the whole box. Here the matter-radiation interaction plays the role of the partition, and when this constraint is released, the system can reach a new equilibrium with entropy given by \eqref{entropie thermodynamique}. Seen in this way, spontaneous emission is analog to a classical Joule Gay-Lussac expansion and the state where the energy quantum is inside the atom (the state when the atom is in its excites state) is just a particular state among others. 

\section{Classical electrodynamics}

\label{classical electrodynamics}

Let consider an oscillating electron hooked to a spring, with pulsation $\omega_0$. If we ignore the electrodynamic laws, the electron movement is just :

\begin{equation}
    \textbf{r}(t) = \textbf{r}_0 e^{i\omega_0 t}
    \label{harmonic oscillator}
\end{equation}

However, from the Maxwell equations, we can recover the Larmor formula giving the radiated power at time $t$ and at distance $r$ from the oscillator \cite{feynman1965feynman} :

\begin{equation}
     P_{ray}(r,t) = \frac{1}{6\pi\epsilon_0c^3}\ddot{d}^2(t-\frac{r}{c})
    \label{puissancerayonnée}
\end{equation}

This involves that the electron mechanical energy decays. We get :

\begin{align}
     \frac{d\langle E \rangle_T}{dt}_{ray} =   \frac{-e^2}{6\pi\epsilon_0c^3}\frac{1}{T}\int_{t}^{t+T}a(t')\frac{dv(t')}{dt'}\mathrm{dt'} \\
     = \frac{+e^2}{6\pi\epsilon_0c^3}\frac{1}{T} \bigg( \int_{t}^{t+T}v(t')\frac{da(t')}{dt'}-[v(t')a(t')]_t^{t+T} \bigg)
    \label{puissancerayonnée2}
\end{align}

where $T = \frac{2 \pi}{\omega_0}$. But $[v(t')a(t')]_t^{t+T} \simeq 0$  because the electron trajectory is almost periodic. Therefore the electron is submitted to a radiative force :

\begin{equation}
     F_{ray}(r,t) = \frac{e^2}{6\pi\epsilon_0c^3}\frac{da(t)}{dt}
    \label{forcerayonnée}
\end{equation}

which in near harmonic regime with pseudo pulsation $\omega_0$ becomes :

\begin{equation}
     F_{ray}(r,t) = -m\frac{v(t)}{\tau}
    \label{forcerayonnée2}
\end{equation}

where :

\begin{equation}
    \frac{1}{\tau} =\frac{e^2\omega_0^2}{6\pi m\epsilon_0c^3}
    \label{constante de temps classique}
\end{equation}

Therefore, at time $t > 0$, the electron trajectory is :

\begin{equation}
    \textbf{r}(t>0) = \textbf{r}_0 e^{-\frac{t}{2\tau}+i\omega_0t}
    \label{amortissement charge}
\end{equation}

Thus, by comparing Eq \eqref{harmonic oscillator} and \eqref{amortissement charge}, we understand that the interaction matter-radius allows new ways of vibrating, because the frequency range broadens. Of course, it is very similar to the spontaneous emission process we studied in the previous sections. But it is also similar to the ball which hits the ground and transforms its mechanical energy into heat. As their total entropy increases because they have new ways of vibrating, the entropy of the electromagnetic field rises because the electron can excite new Fourier modes of the electromagnetic field. Therefore the electromagnetic field entropy increasing should be equal to :

\begin{equation}
    \Delta S = - \sum_{(\textbf{k},s)} p(\textbf{k},s) ln p(\textbf{k},s)
\end{equation}

where $p(\textbf{k},s)$ is the amount of energy of the emitted signal going into the Fourier mode $(\textbf{k},s)$ (divided by the total energy). Of course, the Fourier transform of \eqref{amortissement charge} is easy to calculate, and its amplitude squared gives the energy contained in the Fourier modes. The calculations lead again to a lorentzian distribution with mean value $\omega_0$ and standard deviation $\frac{1}{2\tau}$. Thus, if we enclose our oscillator in a cubic box of volume $V = L^3$, the density of states is $(\frac{L}{2\pi})^3$. Following the same steps and the same approximations as in the previous sections, we get at the end :

\begin{equation}
    \Delta S = ln \frac{V}{V_0}
    \label{entropieclassique}
\end{equation}

With :

\begin{equation}
    V_0 = \frac{6 \pi \tau c^3}{\omega_0^2} 
\end{equation}

Of course, it is totally similar to what we found previously. It's not surprising, the entropy formula \eqref{entropie3DFourier} we found previously does not involve any purely quantum quantity, and it should also apply to the classical case. The oscillator energy splits into many components, because there are many available modes of the electromagnetic field, and this splitting is responsible of the irreversibility. Thanks to \eqref{entropieclassique}, we can interpret the radiation force \eqref{forcerayonnée} as an entropic force. As for the spontaneous emission, the entropy associated to this splitting can be interpreted as the perfect gaz entropy, which enhances a description of the Fourier modes in terms of particules. Therefore, a corpuscular description of the Fourier modes seems to be relevant for matter-radiation interaction.

\section{Conclusion}

We found a quantity for the classical and quantum electromagnetic field which measures the degree of irreversibility of the matter-radiation process. We claimed that we could associate an entropy production to all irreversible processes, and in particular to matter-radiation processes. If we focused on the electromagnetic field here, we can think about irreversibility processes in other field theories, as gravity. Indeed, the gravitational wave emission is analog to the electromagnetic wave emission of the dipole, except that the gravitational wave dipole moment vanishes, and is replaced by the quadrupole moment. However, gravity is a more complex topic than electromagnetism. First, gravitation is not linear. Second, the volume $V$ of the "box" containing the gravitationnal system is itself solution of the field equations. 

\section{Acknowledgements}

I want to thank my friends Martin Caelen, Helmy Chekir, Léonard Ferdinand and Pierre Vallet for their feedback on this work and for having encouraged me to write it down. I want espacially to thank Mark T. Mitchison for his careful reading and his invaluable advice.

\bibliographystyle{apsrev4-1}
\bibliography{biblio}

\end{document}